\documentclass[traditabstract]{aa}
\usepackage{graphicx}
\usepackage{natbib}
\usepackage{txfonts}
\begin{document}

\title{Detection of multiple stellar populations in extragalactic massive clusters with JWST}

 \author{M. Salaris \inst{1}  \and  S. Cassisi \inst{2,3} \and A. Mucciarelli \inst{4,5}  \and D. Nardiello  \inst{6,7}}

   \institute{Astrophysics Research Institute,
     Liverpool John Moores University, IC2, Liverpool Science Park,
     146 Brownlow Hill, Liverpool, L3 5RF, UK 
    \and
      INAF-Osservatorio Astronomico d'Abruzzo, via M. Maggini, sn.
     64100, Teramo, Italy
     \and 
     INFN -  Sezione di Pisa, Largo Pontecorvo 3, 56127 Pisa, Italy
     \and 
 Dipartimento di Fisica e Astronomia, Universit\'a di Bologna, Via Gobetti 93/2, Bologna I-40129, Italy
 \and
INAF-Osservatorio di Astrofisica e Scienza dello Spazio di Bologna, Via Gobetti 93/3, Bologna, II-40129, Italy
\and
Dipartimento di Fisica e Astronomia \lq{Galileo Galilei}\rq, Universit\'a di Padova, Vicolo dell'Osservatorio 3, 35122, Padova, Italy 
\and
INAF-Osservatorio Astronomico di Padova, Vicolo dell'Osservatorio 5, 35122, Padova, Italy      
 }

   \date{Received ; accepted }

 
   \abstract{The discovery both through spectroscopy and photometry 
     of multiple stellar populations (multiple in the sense of non homogeneous chemical abundances,
     with specific patterns of variations of a few light elements) 
     in Galactic globular clusters,
     and in Magellanic Clouds' massive intermediate-age and old clusters, 
     has led to a major change in our views about the formation of these objects. To date, none of the proposed scenarios 
     is able to explain quantitatively all chemical patterns observed in individual clusters, and  
     an extension of the study of multiple populations to resolved extragalactic massive clusters beyond the Magellanic Clouds would
     be welcome, for it would enable 
     to investigate and characterize the presence of multiple populations in different environments and age ranges.
     To this purpose, the \textit{James Webb Space Telescope} can potentially
     play a major role. On the one hand, the \textit{James Webb Space Telescope} promises direct observations of proto-globular cluster candidates at high redshift; 
     on the other hand, it can potentially push to larger distances the sample of resolved clusters with detected multiple populations.
     In this paper we have addressed this second goal. Using theoretical stellar spectra and stellar evolution models, we have investigated 
     the effect of multiple population chemical patterns on synthetic magnitudes in the \textit{James Webb Space Telescope} infrared NIRCam filters. We have identified 
     the colours $(F150W-F460M)$, $(F115W-F460M)$ and pseudocolours $C_{F150W,F460M,F115W}=(F150W-F460M)-(F460M-F115W)$,
     $C_{F150W,F277W,F115W}=(F150W-F277W)-(F277W-F115W)$, as diagnostics able to reveal the presence of multiple populations along the
     red giant branches of old and intermediate age clusters.
     Using the available on-line simulator for the NIRCam detector, we have estimated that multiple populations can be
     potentially detected --depending on the
     exposure times, exact filter combination used, plus the extent of the abundance variations and the cluster [Fe/H]-- 
     out to a distance of $\sim$5~Mpc
     (approximately the distance to the  M83 group).}
\keywords{stars: abundances -- Hertzsprung-Russell and C-M diagrams -- stars: evolution -- globular clusters: general}

\titlerunning{JWST and multiple populations in clusters}
\authorrunning{M. Salaris et al.}
   \maketitle
%

\section{Introduction}\label{intro}

The formation of globular clusters (GCs) is still an open problem \citep[see, e.g.,][for a very recent review]{forbes},
made somewhat more complex by the discovery that GCs do not host single-age, single chemical composition populations, as widely
believed until not so long ago.
Variations of the initial chemical abundances of some light elements in individual Milky Way GCs have been known since about 40 years \cite[see, e.g.,][]{cohen},
however only the more recent advent of high-resolution multi-object spectrographs has solidified this
result \citep[see, e.g.,][and references therein]{carretta09a, carretta09b, gcb}.

In addition to direct spectroscopic measurements, intracluster abundance variations can be detected also
through photometry, thanks to their effect on both stellar effective temperatures, luminosities,
and spectral energy distributions \citep[see, e.g.,][]{sw06, marino, yong, sbordone}. 
The use of appropriate colours and colour combinations (denoted as \textit{pseudocolours})
has indeed allowed to enlarge the sample of clusters, the sample of stars in individual clusters,
and the range of evolutionary phases (including the main sequence, typically too faint to be investigated spectroscopically
with current facilities) where chemical abundance variations have been detected \citep[see, e.g.,][]{sumo, p15, m17, n17}.
Very specifically, photometric studies have for example clearly demonstrated the existence of He abundance variations
in individual clusters, that are associated to the light element variations observed also spectroscopically.

By employing both spectroscopy and photometry, it has been definitely established that individual GCs
host multiple populations (MPs) of stars, characterised by anti-correlations among C, N, O, Na (sometimes also Mg, Al) and He 
\citep[see, e.g., the reviews by][]{gcb, bl18}. Most scenarios for the origin of MPs
\citep[reviewed, e.g., in][]{bl18} invoke subsequent episodes of star formation. Stars with CNONa (and He) abundance ratios similar
to those observed in the halo field are the first objects to form (we will denote them as P1 stars),
while stars enriched in N and Na (and He) and depleted in C and O formed later (we will denote them as P2 stars), 
from freshly synthesised material ejected by some class of massive stars from the first epoch of star formation.
To date, none of the proposed scenarios is able to explain quantitatively all chemical
patterns observed in individual GCs \citep[][]{r15, bl18}. Also, the recent indication 
of He-abundance variations amongst P1 stars in individual clusters \citep[][]{lsb18, He18}
is particularly difficult to accommodate by these standard scenarios.

Additionally, spectroscopic and to a much larger extent photometric studies of small samples of resolved  
extragalactic massive clusters, have shown that the MP phenomenon is not confined to the Milky Way GCs, and that also massive clusters
down to ages of $\sim$2~Gyr do show MPs \citep[see, e.g.,][and references therein]{l14,martocchia,h19,lagioia,nardiello19,mart19}. This latter realization
adds an additional and important piece of information to the MP puzzle, revealing a potential
close connection between the formation of old GCs and young massive clusters. 

It is therefore very important to extend the study of resolved extragalactic massive clusters, to investigate and characterize
the presence of MPs in various environments and age ranges. To this purpose, the \textit{James Webb Space Telescope}
\citep[JWST -- currently scheduled for launch in 2021, see][]{jwst} can potentially
play a very important role. On the one hand, it promises direct observations of proto-GC candidates at high redshift 
\citep[see, e.g.,][]{vanzella, pozzetti}; on the other hand, JWST observations could
potentially push to larger distances the sample of resolved clusters with detected MPs.
Both types of information are necessary to fully understand the MP phenomenon in massive clusters.

The purpose of this paper is to study the effect of the MP chemical patterns on synthetic magnitudes in the JWST infrared NIRCam filter system,
with the aim to identify the
most suitable colours and pseudocolours able to disentangle cluster MPs. We will especially
focus on red giant branch (RGB) stars, because their brightness allow us to maximize the distance out to which MPs can be studied.
Although investigations so far \citep[see, e.g.,][]{sbordone, cmp13, p15} have shown the power of ultraviolet (UV) and near-UV filters to
disentangle photometrically
MPs amongst RGB stars,  
we will show here that also combinations of NIRCam infrared filters can be used for MP detections. 

The paper is structured as follows. Section~\ref{spectral} describes briefly the theoretical 
spectral energy distributions employed in this work and the bolometric correction calculations. Section~\ref{CMD} follows, discussing the 
effect of MPs on the absolute magnitudes of RGB stars in the NIRCam filters, to present a set of colours and pseudocolours
able to reveal the presence of MPs. A final discussion follows in Sect.~\ref{conclusions}.

\section{Spectral energy distributions and bolometric correction calculations}\label{spectral}
\begin{figure}
   \centering
   \includegraphics[width=8.7cm]{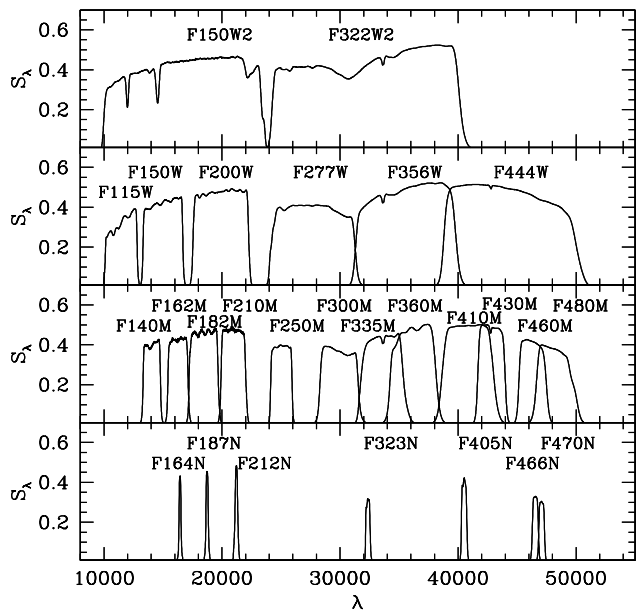}
   \caption{Total throughput as a function of the wavelength $\lambda$ (in \AA) for the 27 NIRcam filters considered in this work.
     From top to bottom, we show extra wide-, wide-, medium- and narrow-band filters.}
         \label{filters}
   \end{figure}

 NIRcam is the primary imager onboard JWST, with a wavelength range
of $6000 < \lambda < 50000 \AA$. We have considered in our analysis 27 out of the 29 NIRcam filters. Their total
(NIRCam + JWST optical telescope element) throughputs are shown in Fig.~\ref{filters}
\footnote{Throughputs taken from \url{https://jwst-docs.stsci.edu/display/JTI/NIRCam + Filters}}. 
The camera has two identical modules, A and B, and the throughputs displayed in Fig.~\ref{filters} are averaged between the modules.
We have discarded the wide band filters $F070W$ and $F090W$ at wavelength between $\sim$6000 and $\sim$10000 \AA, because 
they cover a range already discussed in \citet{sbordone} and \citet{cmp13}, and behave similarly to the Cousins $I$ filter.

As shown in \citet{sw06}, \citet{sbordone} and \citet{cmp13}, the effect of the CNONa(MgAl) abundance 
anticorrelations on the evolutionary properties of low-mass stars (with masses up to about 1.5$M_{\odot}$) 
and the resulting theoretical isochrones is negligible, as long as the CNO sum 
is unchanged compared to the standard $\alpha$-enhanced composition. Available spectroscopic observations
confirm that generally (with only a few exceptions) the CNO sum in P1 and P2 cluster stars is the same within a factor $\sim$2, that is the
typical spectroscopic error bar in these estimates \citep{carretta05}.
This means that studying MPs in the NIRCam filters requires only to assess
the effect of the P2 mixture on the relevant bolometric corrections.

The first step of our analysis was to define the chemical
mixtures for the atmosphere and spectral energy distribution calculations. We have employed the same metal abundance distributions as in
\citet{cmp13} for both P1 and P2 stars, reported in 
Table~\ref{zmixtures}. The P1 mixture is the $\alpha$-enhanced ([$\alpha$/Fe]=0.4) mixture employed in \citet{bastialpha} stellar models, 
whilst the P2 composition has depletions of C, O and Mg by 0.6, 0.8 and 0.3~dex, and enhancements of N, Na and Al by 1.44, 0.8 and 1~dex,
respectively, compared to the P1 abundances. The CNO sum (in both number and mass fractions) is the same in both compositions,
within 0.5\%. On the whole the P2 mixture corresponds to extreme values of the
light element anticorrelations observed in Galactic GCs, as discussed in \citet{cmp13}.

\begin{table}[ht]
\caption{Mass and number fractions (normalized to unity) for the P1 and P2
  metal mixtures considered.}
\label{zmixtures}     
\centering
{\scriptsize                          
\begin{tabular}{lllll} 
\hline
   &  P1   &    &   P2  &          \\
   & Number frac.  & Mass frac.  &  Number frac.  & Mass frac. \\ 
\hline
C  & 0.108211 & 0.076451 &  0.027380 & 0.019250  \\
N  & 0.028462 & 0.023450 &  0.712290 & 0.647230  \\
O  & 0.714945 & 0.672836 &  0.114040 & 0.107660  \\
Ne & 0.071502 & 0.084869 &  0.071970 & 0.085550  \\
Na & 0.000652 & 0.000882 &  0.004137 & 0.005610  \\
Mg & 0.029125 & 0.041639 &  0.014660 & 0.020990  \\
Al & 0.000900 & 0.001428 &  0.000906 & 0.014400  \\
Si & 0.021591 & 0.035669 &  0.021730 & 0.035960  \\
P  & 0.000086 & 0.000157 &  0.000087 & 0.000158  \\
S  & 0.010575 & 0.019942 &  0.010640 & 0.020100  \\
Cl & 0.000096 & 0.000201 &  0.000097 & 0.000203  \\
Ar & 0.001010 & 0.002373 &  0.001017 & 0.002390  \\
K  & 0.000040 & 0.000092 &  0.000040 & 0.000093  \\
Ca & 0.002210 & 0.005209 &  0.002244 & 0.005251  \\
Ti & 0.000137 & 0.000387 &  0.000138 & 0.000390  \\
Cr & 0.000145 & 0.000443 &  0.000146 & 0.000466  \\
Mn & 0.000075 & 0.000242 &  0.000075 & 0.000244  \\
Fe & 0.009642 & 0.031675 &  0.009705 & 0.031930  \\
Ni & 0.000595 & 0.002056 &  0.000599 & 0.002073  \\
\hline
\end{tabular}}
\end{table}

We have then considered a set of three 12~Gyr reference $\alpha$-enhanced isochrones from the BaSTI
database\footnote{\url{http://basti.oa-abruzzo.inaf.it/index.html}.} \citep{bastialpha}, for [Fe/H]=$-$1.62, $Y$=0.246,
[Fe/H]=$-$0.70, $Y$=0.256 and [Fe/H]=$-$0.70, $Y$=0.300, this latter to include the effect of He-enhancements in P2 stars.
Along these three isochrones we selected eight key points that cover almost the full range of RGB effective
temperature $T_{\mathrm eff}$ and luminosities, and
reach down well below the main sequence turn off.
For each of these points we have then calculated appropriate model atmospheres and synthetic spectra
for each of the metal mixture/He mass fraction pairs described before.
The parameters of the model atmosphere calculations are reported in Table~\ref{parameters}.

\begin{table*}[ht]
\caption{Effective temperature, surface gravity and metallicity of the calculated synthetic spectra.}
\label{parameters}     
\centering                          
\begin{tabular}{cccccc}       
\hline                
$T_{\mathrm eff}$ & $\log g$ & $T_{\mathrm eff}$ & $\log g$ & $T_{\mathrm eff}$ & $\log g$ \\
\hline
 [Fe/H]=$-$0.7 & & [Fe/H]=$-$1.62 & & [Fe/H]=$-$0.7,$Y$=0.3 &  \\
\hline
4501  &  4.67 &  4621 &  4.77 & 4704 &  4.67 \\
5550  &  4.53 &  6131 &  4.50 & 5654 &  4.53 \\
5998  &  4.24 &  6490 &  4.22 & 6041 &  4.24 \\
5502  &  3.91 &  5854 &  3.78 & 5566 &  3.91 \\
5050  &  3.36 &  5312 &  3.21 & 5101 &  3.36 \\ 
4551  &  2.03 &  4892 &  2.06 & 4580 &  2.03 \\ 
4052  &  1.12 &  4476 &  1.20 & 4078 &  1.12 \\
3500  &  0.11 &  4100 &  0.50 & 3500 &  0.08 \\
\hline   
\end{tabular}
\end{table*}

Model atmospheres and synthetic spectra have been computed as in \citet{cmp13}.
For each set of parameters in Table~\ref{parameters}, a plane-parallel, local thermodynamical equilibrium, 1-dimensional model atmosphere
has been calculated with the ATLAS12 \citep{kurucz05, sbordone07} code, that employs the
opacity sampling method to compute model atmospheres with an
arbitrary chemical composition.
Synthetic spectra have been then calculated with the SYNTHE \citep{kurucz05} code
in the spectral range 8000 - 52000 \AA\, including all molecular 
and atomic lines provided in the Kurucz/Castelli compilation
\footnote{\url{http://wwwuser.oats.inaf.it/castelli/linelists.html}}, as well as
all predicted levels usually adopted in the calculation of
colour indices and flux distributions.
The TiO transitions \citep{schwenke98} have been included only for the models with $T_{\mathrm eff} <$4100~K.
We adopted for each synthetic spectrum the new release of the H$_2$O linelist
by \citet{ps97} provided by R.L.Kurucz \footnote{\url{http://kurucz.harvard.edu/molecules/h2o/}}.

\begin{figure}
   \centering
   \includegraphics[width=8.7cm]{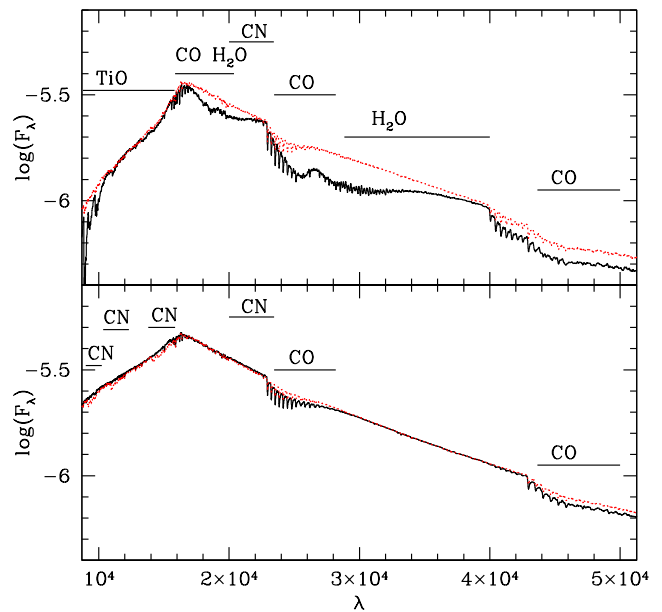}
   \caption{SED for P1 (solid lines) and P2 (dotted lines) [Fe/H]=$-$0.7 RGB models with $T_{\mathrm eff}$=4052~K,
     $\log g$=1.12 (lower panel) and $T_{\mathrm eff}$=3500~K,
     $\log g$=0.11 (upper panel). The vertical axis displays the logarithm of the flux in $erg/sec/cm{^2}/Hz/ster$, whilst the horizontal axis
     shows the wavelength in \AA.
     The labels denote the main molecules contributing to the SED in those wavelength ranges
     where P1 and P2 results are different.}
         \label{spectra}
   \end{figure}

Figure~\ref{spectra} compares P1 and P2 spectral energy distributions (SEDs) for [Fe/H]=$-$0.7 models with $T_{\mathrm eff}$=4052~K,
$\log g$=1.12, and $T_{\mathrm eff}$=3500~K,
$\log g$=0.11, respectively. There are clearly very specific wavelength ranges where the spectra differ due to the
different metal compositions. The molecules that affect the SED in those wavelength regions are labelled.

Starting from these three sets of theoretical SEDs, whose parameters are given in Table~\ref{parameters}, we have calculated
bolometric corrections (BCs) for the NIRCam filters in the VEGAmag system
following \citet{g02}, setting the solar bolometric magnitude to 4.74 according to the IAU
recommendations\footnote{\url{https://www.iau.org/static/resolutions/IAU2015_English.pdf}}.

\section{Colour- and pseudocolour-magnitude diagrams}\label{CMD}

\begin{figure}
   \centering
    \includegraphics[width=8.7cm]{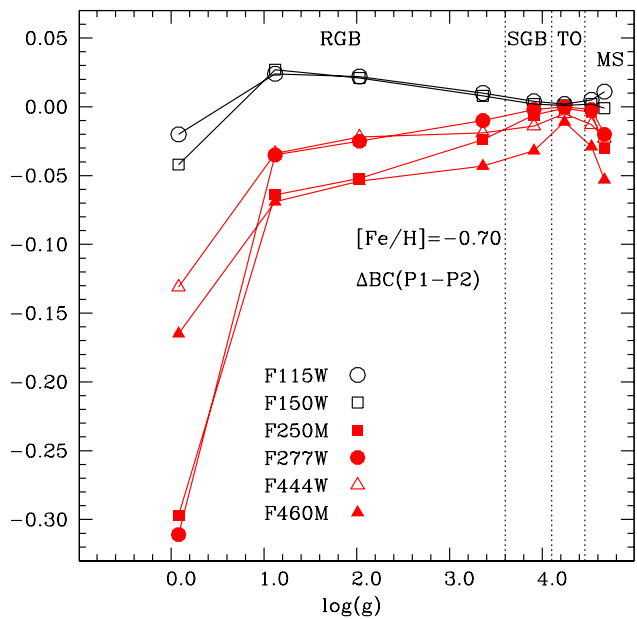}
    \caption{Difference ($\Delta$(BC)) between BCs calculated for P1 and P2
      compositions (P1-P2), at selected points along a 12~Gyr isochrone with [Fe/H]=$-$0.7
      and $Y$=0.256 (see Table~\ref{parameters}). $\Delta$(BC) is plotted as a function of the surface gravity, that decreases steadily 
      from the lower main sequence to the tip of the RGB. We also label the evolutionary
      phases --main sequence (MS), turn off (TO), subgiant branch (SGB) and RGB--  corresponding to the selected isochrone points.
      Only filters showing the largest positive and negative differences are displayed
      (see text for details).}
    \label{figdiffbc}
  \end{figure}

\begin{figure}
   \centering
    \includegraphics[width=8.7cm]{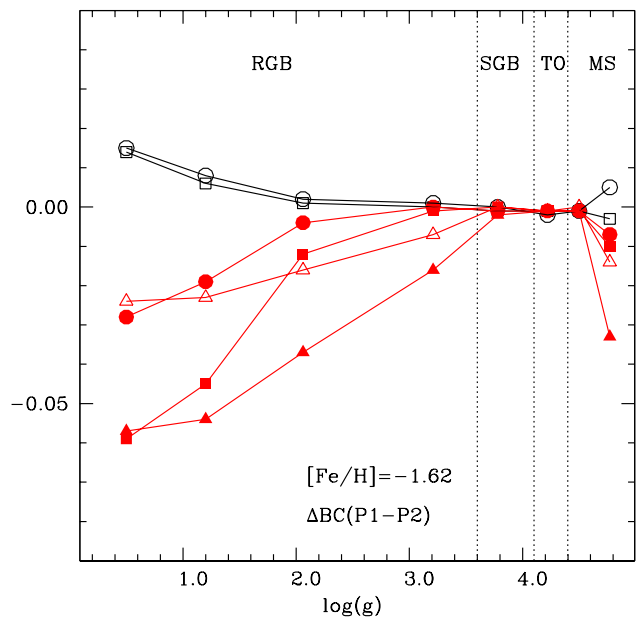}
    \caption{As Fig.~\ref{figdiffbc} but for [Fe/H]=$-$1.62.}
    \label{figdiffbcb}
\end{figure}

A comparison of the BCs for P1 and P2 models provides the necessary guideline
to assess whether NIRCam filters can identify the presence of multiple populations in massive clusters.
Figures~\ref{figdiffbc} and ~\ref{figdiffbcb} display the difference between P1 and P2 models (P1-P2) at [Fe/H]=$-$0.7 and [Fe/H]=$-$1.62. Results for the case of [Fe/H]=$-$0.70 and $Y$=0.30 are within at most $\sim$0.01~mag of the results for [Fe/H]=$-$0.7 and normal $Y$.

In general, the differences of the BCs ($\Delta$(BC)) for all filters
have a minimum around the main sequence turn off (located at log($g$)$\sim$4.0), and tend to increase moving along the RGB or going
down along the main sequence. This is clearly due to a dependence of $\Delta$(BC) on $T_{\mathrm eff}$, with $\Delta$(BC) increasing
with decreasing $T_{\mathrm eff}$ \citep[see also][for a similar result with Johnson-Cousins and Str\"omgren filters]{sbordone, cmp13}.
The exact values of $\Delta$(BC) depend on the metallicity of the models, decreasing in absolute values when
[Fe/H] decreases.

Focusing now on the RGB, the filters $F115W$ and $F150W$ show the largest positive $\Delta$(BC), whereas $F250M$, $F277W$, $F444W$
and $F460M$ display the largest negative variations. The narrow band filters $F466N$ and $F470N$
(not displayed in Figs.~\ref{figdiffbc} and ~\ref{figdiffbcb}) follow exactly the same results as $F460M$.
Notice that below $\sim$4000~K (log($g$)$\sim$1) at [Fe/H]=$-$0.7, $F250M$ and $F277W$ display a huge sudden decrease of $\Delta$(BC). 

\begin{figure}
   \centering
   \includegraphics[width=8.7cm]{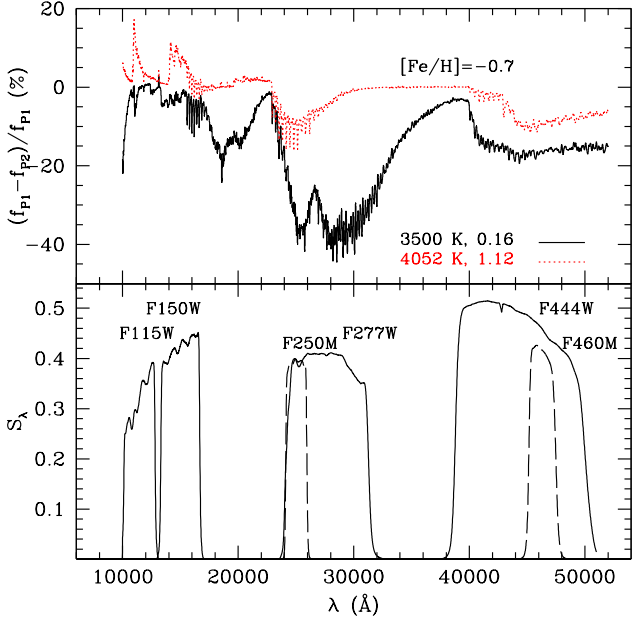}
   \caption{{\sl Upper panel}: Percentage flux difference between the SED of P1 ($f_{P1}$) and P2 ($f_{P2}$) models  
     with the labelled values of $T_{\mathrm eff}$ and $\log g$, as a function of the wavelength $\lambda$ (in \AA), 
     in the range covered by the NIRcam filters discussed here. {\sl Lower panel}: Total throughput of the wide and medium-band
     filters that cover the wavelength range showing the largest differences in the SED.}
         \label{filtdiff}
   \end{figure}

Figure~\ref{filtdiff} explains why the above-mentioned filters are the ones most affected by the P2 chemical composition.
This figure shows the percentage difference of the SED between P1 and P2 models at the two coolest point along the [Fe/H]=$-$0.7 RGB,
together with the passbands of the filters displayed in Fig.~\ref{figdiffbc}.

We consider first the spectra for $T_{\mathrm eff}=$4052~K. SED differences are localized at specific wavelength ranges,
and reflect the effect of the molecular abundances labelled in Fig.~\ref{spectra}.
The $F115W$ and $F150W$ filters are sensitive to variations of CN, that cause a decrease of the P2 flux compared to the P1
counterpart -- the increase of N, much less abundant than C in the P1 composition, dominates over the decrease of C 
and causes an increase the strength of the CN molecular lines in the P2 SED-- whereas
$F250M$, $F277W$, $F444W$ and $F460M$ are sensitive to variations of CO, that increase the P2 flux compared to the
P1 values (both C and O abundances 
in the P2 composition are decreased). This behaviour is typical of all SEDs along the RGB, apart from the coolest point
at [Fe/H]=$-$0.7 ($T_{\mathrm eff}=$3500~K).

Indeed, at $T_{\mathrm eff}=$3500~K the situation is somewhat different, for 
in the wavelength range of $F115W$ and $F150W$ it is now the TiO absorption that starts to dominate.
Due TiO variations $\Delta$(BC) values for these two filters change sign compared to higher $T_{\mathrm eff}$ (see Fig.~\ref{figdiffbc}), but
with values still close to zero.

For the $F444W$ and $F460M$ filters the CO absorption still dominates.
In the wavelength regime of $F250M$ and $F277W$, the ${\rm H_2O}$ absorption
becomes important above $\sim 26000$ \AA, whilst between $\sim$22000 and $\sim$26000 \AA\, it is still mainly CO.
Between $\sim$16000 and $\sim$22000 \AA\, variations of both CO and ${\rm H_2O}$ cause the observed flux differences, that were essentially zero at $T_{\mathrm eff}=$4052~K.
At [Fe/H]=$-$1.62 the RGB model $T_{\mathrm eff}$ does never become low enough to see the effect on the BCs of TiO and ${\rm H_2O}$ variations. 

We emphasize here that these $\Delta$(BC) values are preserved also when calculated for younger RGB isochrones. Considering for example
a 2~Gyr, [Fe/H]=$-$0.7 isochrone, for a given $T_{\mathrm eff}$ on the RGB the surface gravity is changed by at most
0.1-0.2~dex compared to the $\log g$ values at the same $T_{\mathrm eff}$ on a 12~Gyr isochrone.
We have tested that these small changes of $\log g$ do not affect the $\Delta$(BC) results
for the various NIRCam filters.
   
\begin{figure}
   \centering
   \includegraphics[width=8.7cm]{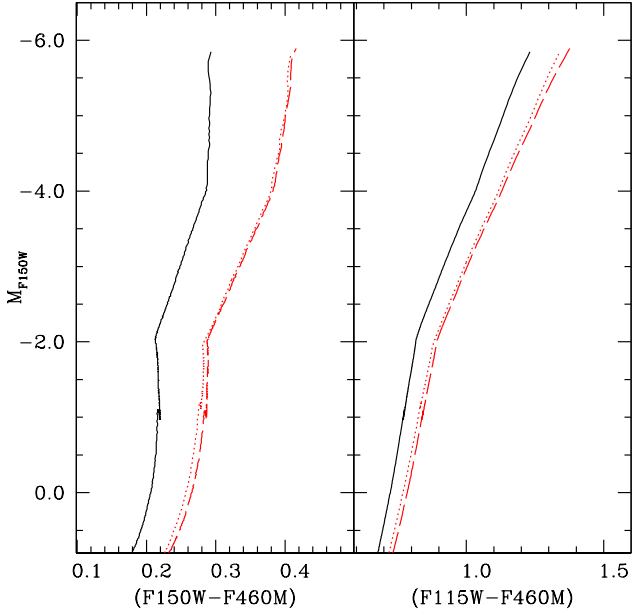}
   \caption{$M_{F150W}$-$(F150W-F460M)$ and $M_{F150W}$-$(F115W-F460M)$ diagram for 12~Gyr, [Fe/H]=$-$0.7, P1 (solid line) and P2 --
     with normal (dashed line) and enhanced (dotted line) initial He --
   isochrones.}
         \label{diagrams}
   \end{figure}

From the previous discussion we can conclude that colour-magnitude-diagrams (CMDs) like $M_{F150W}$-$(F150W-F460M)$, or
$M_{F115W}$-$(F115W-F460M)$ shown in Fig.~\ref{diagrams} -- where we applied the calculated BCs to the
reference 12~Gyr [Fe/H]=$-$0.7, $Y$=0.256 isochrone -- can in principle reveal the presence of multiple populations along
the cluster RGB. The P2 RGB is redder than the P1 counterpart, and  
the maximum colour separation with the chosen P2 pattern --that is at the upper limit of the observed
range of abundance anticorrelations in Galactic GCs-- is of about 0.10~mag along the upper RGB.
At [Fe/H]=$-$1.62 this difference decreases by $\sim$0.01-0.02~mag.
If the P2 composition has enhanced He ($Y$=0.30), the separation between P1 and P2 RGBs is reduced by $~\sim$0.01~mag. 

It is possible in principle to consider also CMDs like $M_{F150W}$-$(F150W-F277W)$ and
$M_{F150W}$-$(F150W-F250W)$ to maximize 
the separation of P1 and P2 only at magnitudes close to the RGB tip
and at high metallicities (see results in Figs.~\ref{figdiffbc} and ~\ref{figdiffbcb}).

\begin{figure}
   \centering
   \includegraphics[width=8.7cm]{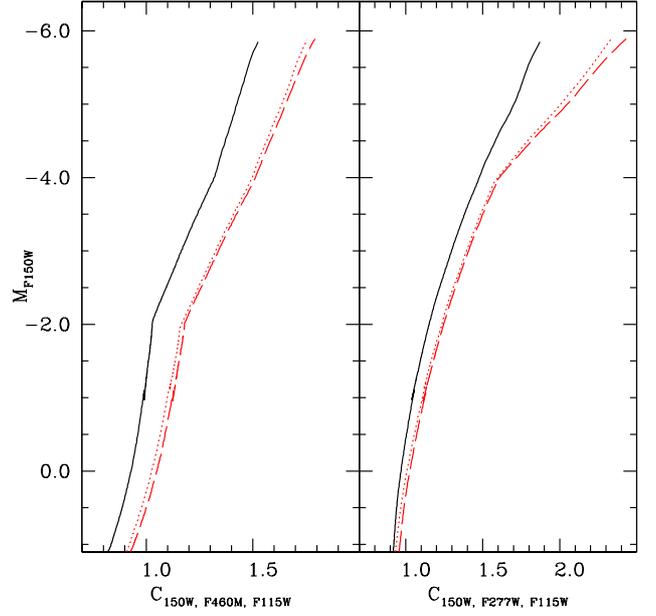}
   \caption{As Fig.~\ref{diagrams} but for the $M_{F150W}$-$C_{F150W,F460M,F115W}$ and $M_{F150W}$-$C_{F150W,F277W,F115W}$ diagrams.}
         \label{pseudocolour}
   \end{figure}

To maximize the separation of P1 and P2 RGB tracks it is convenient to define -- in a similar way as \citet{mmp}, \citet{p15}--
a pseudocolour, that is the difference of two colours that change with opposite signs when going from P1 to P2 composition. 
We have defined here the pseudocolour $C_{F150W,F460M,F115W}=(F150W-F460M)-(F460M-F115W)$, and we show in Fig.~\ref{pseudocolour} the
$M_{F150W}$-$C_{F150W,F460M,F115W}$ diagram for the same [Fe/H]=$-$0.7 isochrones of Fig.~\ref{diagrams}.
The separation in $C_{F150W,F460M,F115W}$ between P1 and P2 isochrones is of the order of $\sim$0.1~mag along the lower RGB,
increasing to $\sim$0.2~mag along the upper RGB, with a very small
decrease if the P2 composition is He-enhanced. At [Fe/H]=$-$1.62 these figures are decreased by about 0.02-0.03~mag.

The alternative pseudocolour $C_{F150W,F277W,F115W}=(F150W-F277W)-(F277W-F115W)$ would enhance the separation of P1 and P2 RGBs compared to
$C_{F150W,F460M,F115W}$, with differences  
up to $\sim$0.5~mag, but from only about 1~mag below the RGB tip in $M_{F150W}$, and for metal rich compositions. 

The effect of extinction in these photometric bands is, not suprisingly, small. We have determined the extinction
relationships $A_{\lambda}/A_V$ for what we considered to be the best choice $F115W$, $F150W$ and $F460M$ filters discussed here,
employing \cite{ccm} extinction law with $R_V=A_V/E(B-V)$=3.1 and the formalism by \citet{g02}.  We found
$A_{F115W}/A_V$=0.32,  $A_{F150W}/A_V$=0.21 and $A_{F460M}/A_V$=0.05, implying for the reddening affecting the 
pseudocolour $C_{F150W,F460M,F115W}$ a relationship $E(C_{F150W,F460M,F115W})=0.43 A_V$.
If we consider the pseudocolour $C_{F150W,F277W,F115W}$ , we find that $A_{F277W}/A_V$=0.09 and $E(C_{F150W,F277W,F115W})=0.35 A_V$.
To consider the case of extragalactic extinction laws with $R_V$ different from 3.1, we assumed $R_V$=5 as an example.
We found in this case $A_{F115W}/A_V$=0.38,  $A_{F150W}/A_V$=0.25, $A_{F460M}/A_V$=0.05, $A_{F277W}/A_V$=0.10,  
$E(C_{F150W,F460M,F115W})=0.53 A_V$ and $E(C_{F150W,F277W,F115W})=0.43 A_V$, values that are only marginally different from the results for $R_V=3.1$.

\section{Discussion}
\label{conclusions}

In this paper we have studied the effect of MP chemical patterns on synthetic magnitudes in the JWST NIRCam filter system, 
identifying colours and pseudocolours able to disentangle MPs among clusters' RGB stars.
We found that colours like $(F150W-F460M)$ and $(F115W-F460M)$, or the pseudocolour
$C_{F150W,F460M,F115W}$ are well suited to separate MPs over a reasonably large RGB temperature range, due to their sensitivity to variations of
CN, CO and, with increasing metallicity, also TiO and ${\rm H_2O}$ molecular abundances.

With P2 light element abundance spreads typical of the extreme patterns observed in Galactic GCs, 
$C_{F150W,F460M,F115W}$ is predicted to display at [Fe/H]=$-$0.7 a range (at fixed luminosity) of the order of $\sim$0.1~mag along the lower RGB,
increasing to $\sim$0.2~mag along the upper RGB. In case of the $(F150W-F460M)$ and $(F115W-F460M)$ colours, the expected range 
is of about 0.10~mag along the upper RGB.
At a lower [Fe/H]=$-$1.62 these figures are reduced by 0.02-0.03~mag.
Smaller MP abundance anticorrelations' ranges will of course cause smaller colour and pseudocolour spreads,
and the detection of MPs in extragalactic clusters
using NIRCAM filters will depend on the actual photometric errors plus
the range of cluster abundance anticorrelations in the target cluster, and the cluster [Fe/H].

It is interesting and useful to have a general idea of 
the maximum distance out to which it will be possible to disentangle MPs with JWST, at least for clusters with sizable abundance spreads. 
In the following experiment we require 0.01~mag photometric errors down to a couple of magnitudes below the RGB tip
in the relevant NIRCAM filters.
These photometric errors would enable us to definitely disentangle MPs in clusters at both [Fe/H]=$-$1.6 and $-$0.7, with 
CNONa abundance variations even lower than the maximum amplitudes observed in Galactic GCs.
 
We made use of the on-line
JWST simulator for the NIRCam detector\footnote{\url{https://jwst.etc.stsci.edu}}, and
considered a RGB star of spectral type K0III 
--taken as representative of bright GC RGB objects-- using the
corresponding SED from the Phoenix stellar model atmosphere library \citep{husser}.
As a reasonable exposure time, we employed a
set of 5 exposures, each one by 3 integrations, that are
obtained with 8 groups (for a total of 15 integrations and total exposure time of about 
24000\,s)\footnote{See \url{https://jwst-docs.stsci.edu/display/JPPOM/Exposure+Timing} for details about exposure timings.}.

To simulate the blending effects, we computed the average separation of
RGB stars located at a distance $>1.7$\,arcmin from the centre of the
GC 47\,Tuc - taken as representative of a typical crowding condition among GCs.
We reported this separation to different distances, and for each distance we determined
the corresponding angular separation, denoted by $s$. We considered then two representative RGB stars
separated by $s$, and performed aperture photometry with radius $s$.
We calculated the
S/N of a single RGB star, considering the worst case that 50\,\% of
the flux of the neighbouring object falls within the aperture. 

With these assumptions, we found that at a distance of $\sim$1.4\,Mpc \citep[beyond the Andromeda Group, see, e.g.,][]{kar}
we can still achieve the required 0.01~mag photometric error down to $\sim$2~mag below the RGB tip
in the filters $F115W$, $F150W$ and $F460M$.

The F460M filter, that has a narrower passband compared to $F115W$ and $F150W$, does essentially set this distance for a fixed photometric error.
We could in principle use the wider $F444W$ filter that is sensitive to the same molecular features as $F460M$ (see Figs.~\ref{spectra} and \ref{filtdiff}),
but the price to pay is a reduction of the predicted sensitivity to MPs of the corresponding colours (see Figs.~\ref{figdiffbc} and ~\ref{figdiffbcb}).
By employing the pseudocolour based on $F277W$ ($C_{F150W,F277W,F115W}$, see Fig.~\ref{pseudocolour}) we would be able to disentangle MPs with a better or comparable
sensitivity only down to $\sim$1~mag below the RGB tip.
Nevertheless, with these alternative filter combinations and the same observational setup, the JWST simulator suggests that 
we could achieve 0.01~mag photometric errors two magnitudes below the RGB tip in all photometric filters at a distance of $\sim$5\,Mpc \citep[roughly the 
distance to the M83 Group and the Canes Venatici I Cloud, see e.g.,][]{kar}.

To maximize the distance with the preferred colour/pseudocolour combinations involving the filter $F460M$, we studied an extreme case
for a set of 10 exposures, each one by 5 integrations, obtained with 8 groups (total exposure time $\sim$ 80000\,s). In this case
we can achieve a 0.01~mag photometric error down to $\sim$2~mag below the RGB tip at a distance of $\sim$2.3~Mpc.

The standardization of real JWST data no doubt will turn out to be to some degree different from 
what we have employed in this analysis. Still, these results make it possible
to obtain a first estimate of the appearance of MPs through the eye of JWST, and to assist with the planning of
future observations when the telescope will be operational.

\begin{acknowledgements}
We thank our anonymous referee for comments that helped improve the presentation of our results.
SC acknowledges support from Premiale INAF MITiC, from INFN (Iniziativa specifica TAsP), and 
grant AYA2013-42781P from the Ministry of Economy and Competitiveness of Spain.
DN acknowledges partial support by the Universit\`a degli Studi
di Padova, Progetto di Ateneo BIRD178590.

\end{acknowledgements}

\bibliographystyle{aa}

\end{document}